\documentclass[manuscript]{emulateapj}

\slugcomment{Submitted to ApJ Letters}

\shorttitle{The Frequency of CEMP Stars}
\shortauthors{Lucatello et al.}

\begin{document}

\title{The Frequency of Carbon-Enhanced Metal-Poor Stars in the Galaxy from the
HERES sample}

\author{Sara Lucatello}
\affil{Osservatorio Astronomico di Padova, Vicolo dell'Osservatorio 5, 35122, Padova, Italy}
\email{lucatello@pd.astro.it}

\author{Timothy C. Beers}
%, Brian Marsteller, Thirupathi Sivarani, Young Sun Lee}
\affil{Department of Physics \& Astronomy, CSCE: Center for the Study of Cosmic Evolution,
and JINA: Joint Institute for Nuclear Astrophysics, Michigan State University, East Lansing, MI 48824, USA}
\email{beers@pa.msu.edu}
%,marsteller@pa.msu.edu, thirupathi@pa.msu.edu, leeyou25@msu.edu}

\author{Norbert Christlieb}
\affil{Hamburger Sternwarte, University of Hamburg, Gojenbergsweg 112, D-21029, Hamburg, Germany;\\
Department of Astronomy and Space Physics, Uppsala University, Box 515, 751-20 Uppsala, Sweden}
\email{nchristlieb@hs.uni-hamburg.de}

\author{Paul S. Barklem}
\affil{Department of Astronomy and Space Physics, Uppsala University, Box 515, 751-20 Uppsala, Sweden}
\email{barklem@astro.uu.se}

\author{Silvia Rossi}
\affil{Instituto de Astronomia,  Geof\'{i}sica e Ci\^{e}ncias Atmosf\'{e}ricas, Departamento de Astronomia, 
Universidade de S\~{a}o Paulo, \\ 
Rua do Mat\~{a}o  1226, 05508-900 S\~{a}o Paulo, Brazil}
\email{rossi@astro.iag.usp.br}

\author{Brian Marsteller, Thirupathi Sivarani, Young Sun Lee}
\affil{Department of Physics \& Astronomy, CSCE: Center for the Study of Cosmic Evolution,
and JINA: Joint Institute for Nuclear Astrophysics, Michigan State University, East Lansing, MI 48824, USA}
\email{marsteller@pa.msu.edu, thirupathi@pa.msu.edu, leeyou25@msu.edu} 

\begin{abstract}

We estimate the frequency of Carbon-Enhanced Metal-Poor (CEMP) stars among
very metal-poor stars, based on an analysis of 349 stars with available
high-resolution spectra observed as part of the Hamburg/ESO R-process Enhanced
Star (HERES) survey. We obtain that {\it a lower limit of} 21 $\pm$ 2\% of stars
with [Fe/H] $\leq -2.0$ exhibit [C/Fe] $\geq +1.0$. These fractions are higher than
have been reported by recent examinations of this question, based on
substantially smaller samples of stars. We discuss the source of this difference
and suggest that in order to take into account effects
that result in a decrease of surface carbon abundance with advancing evolution,
a definition of CEMP stars based on a [C/Fe] cutoff that varies as a function of
luminosity is more appropriate.
We discuss the likely
occurrence of dilution and mixing for many CEMP stars, which, if properly accounted
for, would increase this fraction still further.

\end{abstract}


\keywords{stars: abundances---stars: carbon---stars: Population II---Galaxy: halo}

\section{Introduction}

One of the most interesting results from modern surveys for metal-poor stars,
such as the HK survey of Beers and colleagues
\citep{Beers85,Beers92,Beers99} and the Hamburg/ESO Survey (HES) of
Christlieb and collaborators \citep{Christlieb03}, concerns the apparently
large fraction of Carbon-Enhanced Metal-Poor (CEMP) stars identified at
low metallicity. The fraction of CEMP stars among samples of stars with [Fe/H]$
\leq -$2.0 has been reported to be as high as $\sim $20-25\% \citep{Marsteller05},
which is significantly larger than the fraction observed at higher
metallicities. Beers \& Christlieb (2005) have pointed out that this fraction
appears to increase further with decreasing metallicity, reaching $\sim $40\%
for stars with [Fe/H] $< -3.5$. As discussed in \citet{Abia2001} and
\citet{Lucatello05}, this increase in frequency has an immediate consequence on
the nature of the IMF in the early Galaxy, providing evidence for its being
shifted toward higher masses with respect to the present-day IMF.

Recently, this result has been challenged by \citet{Cohen05}, who reported on
the frequency of C-rich stars in a sample of 122 metal-poor giants from the HES
(55 of which have high resolution spectroscopic data),
finding a value of 7.4 $\pm$ 2.9\% for C-stars (defined in that paper as those
objects whose spectra show visible C$_2$ bands), or 14 $\pm$ 4\% when including
in the sample the C-enriched ([C/Fe] $ \geq +$1.0) stars for which no C$_2$ band
was detectable.
\citet{Frebel06}, on the basis of an itermediate 
resolution analysis of a  sample of 147 local stars, find an overall 
CEMP fraction of 9$\pm$2\% among giants, which rises to more than 20\% with 
increasing distance from the Galactic plane.

In this {\it Letter} we consider this question anew, making use of the data
obtained by the Hamburg/ESO R-process Enhanced Star (HERES) survey (Christlieb
et al. 2004, Paper I; Barklem et al. 2005, Paper II), which presently provides
the largest sample of high-resolution spectra available for Very Metal-Poor
stars (VMP; [Fe/H] $\le -2.0$ according to the nomenclature of Beers \&
Christlieb 2005). It is important to note that the selection criteria for HERES
stars was specifically set to minimize biases in the abundances of elements
other than iron, including carbon, hence this sample is well-suited for the
determination of the relative fractions of stars with various abundance
signatures as a function of metallicity.

The stars in the HERES sample that exhibit very strong molecular features
 of carbon (either due to the high C abundances, low stellar
temperatures, or both) are a considerable challenge for the automated 
line profile method adopted in Paper II, hence the latter did not report on the
abundance analyses for 72
stars with available HERES spectra. Here we report estimates of the abundances
of [Fe/H], C, and N for 94 HERES stars -- the 72 stars not presented in Paper II
plus 22 previously analyzed stars having moderately strong molecular features. A
full analysis of additional elemental species for these stars will be reported
in a future paper (Lucatello et al., in preparation)

In \S 2 we summarize the HERES observations and our abundance analysis.  In \S 3
we present our results for the fraction of CEMP stars among the HERES sample,
and describe an alternative luminosity-dependent definition for C-enhanced stars.  
Our conclusions are presented in \S 4.

\section{Observations and Analysis}

Details of the observations and data reduction procedures are
discussed in Papers I and II. Our analysis procedures are described below. 

%\section{Analysis}

\subsection{Atmospheric Parameters}

The quality of the HERES data is generally not sufficiently high for the
derivation of precise spectroscopic temperature estimates. Moreover, because of the
forest of molecular lines (CH, C$_2$, and CN) present in the spectra of many of
the C-rich stars, the measurement of a set of clean Fe lines to use for excitation
equilibrium is often impossible. In Paper II we adopted a temperature scale
based on several broadband optical and infrared colors both from our own
photometry \citep{Beers06} and from the 2MASS Point Source
Catalog \citep{Skrutskie06}. However, the molecular bands present in the C-rich HERES stars
can lead to very discrepant color-based temperatures, depending on the colors that are
used. Given the strong absorption of the CH G band in essentially all of the
C-rich HERES stars, the $B$ passband is particularly affected. For this reason,
and also because of uncertainties in the colors adopted for the selection of HERES
candidates (originally estimated from the HES plates), the $ B-V > $ 0.5
selection criterion for HERES still includes a number of stars that are, in
fact, warmer than intended.

We adopt $V-K$ as the preferred temperature indicator (both $V$ and $K$ passbands 
are only marginally affected by the presence of the strong molecular lines in
the temperature range considered herein, T$_{\rm eff}> $4200\,K), and the
\citet{Alonso99} temperature scale for our analysis. The temperatures derived in
this way trace extremely well those derived by fitting H$_\alpha$ and H$_\beta$
profiles \citep[see][]{Lucatello04}. The adopted reddening estimate is taken
from \citet{Schlegel98} for small to moderate reddening values (E($B-V$)$ < $0.1).
For reddening estimates from Schlegel et al. that exceed this value, we employ
the empirical correction provided by \citet{Bonifacio00}.

Surface gravity values are derived
from the Y$^2$ isochrones \citep{Kim02}, assuming an age of 12\,Gyr, an 
$\alpha$-element enhancement of [$\alpha$/Fe]=$+$0.3.
We assumed a first-pass gravity derived using the isochrone with a metallicity of 
[Fe/H]=$-$2.5 for all of the stars, then after performing the abundance analysis
a new gravity was derived from the isochrone of appropriate metallicity.
The procedure was repeated until convergence for each one of the stars.
\citet{Kurucz03} solar scaled model atmospheres were adopted for our analysis, in order
to be consistent with previous work on C-rich stars.
  The impact of appropriately C- (and N-)
enhanced models will be considered in a future paper. 

\subsection{Chemical Abundance Measurements}

We measured the Fe abundances for our program sample from the equivalent widths
of reasonably clean Fe {\sc I} and {\sc II} lines. The number of clean Fe~I
lines varies between a low of 13 (for HE~1413-1954, which is a hot, metal-poor star 
with a fairly low signal-to-noise spectrum) to about 50.
We find that the derived metallicities for the C-rich 
stars are typically higher than reported in the HES;
a similar result was reported also in \citep{Cohen05}.
The difference is probably due to the adoption of a warmer temperature scale,
based on the $V-K$ color, rather than $B-V$ which can be heavily 
affected by the CH molecular lines, as discussed in, e.g. \citet{Lucatello04}.

Carbon abundances are derived from the synthesis of the CH G-band at
$\sim$4300\,{\AA}; N abundances are obtained from the violet CN system at
$\sim$3800\,\AA. The O abundance, for lack of better information (no O 
features are present in our wavelength range), was adopted to be [O/Fe]=$+$0.4\footnote{Any 
choice of O abundance within the usual range for halo stars, [O/Fe]=0.3-0.6 affects 
the derived C abundances negligibly. In fact, given that the typical 
[C/Fe] in the present sample is $\sim+$1, even assuming that all the O present in 
the star forms CO, the amount of C locked away in CO is negligible with respect to 
to the total C abundance.}.
Details concerning the adopted atomic and molecular parameters
for the spectral analyses and syntheses are given in \citet{Lucatello03}.

\section{Results and Interpretation}

\subsection{The CEMP Fraction}

\begin{figure}
\plotone{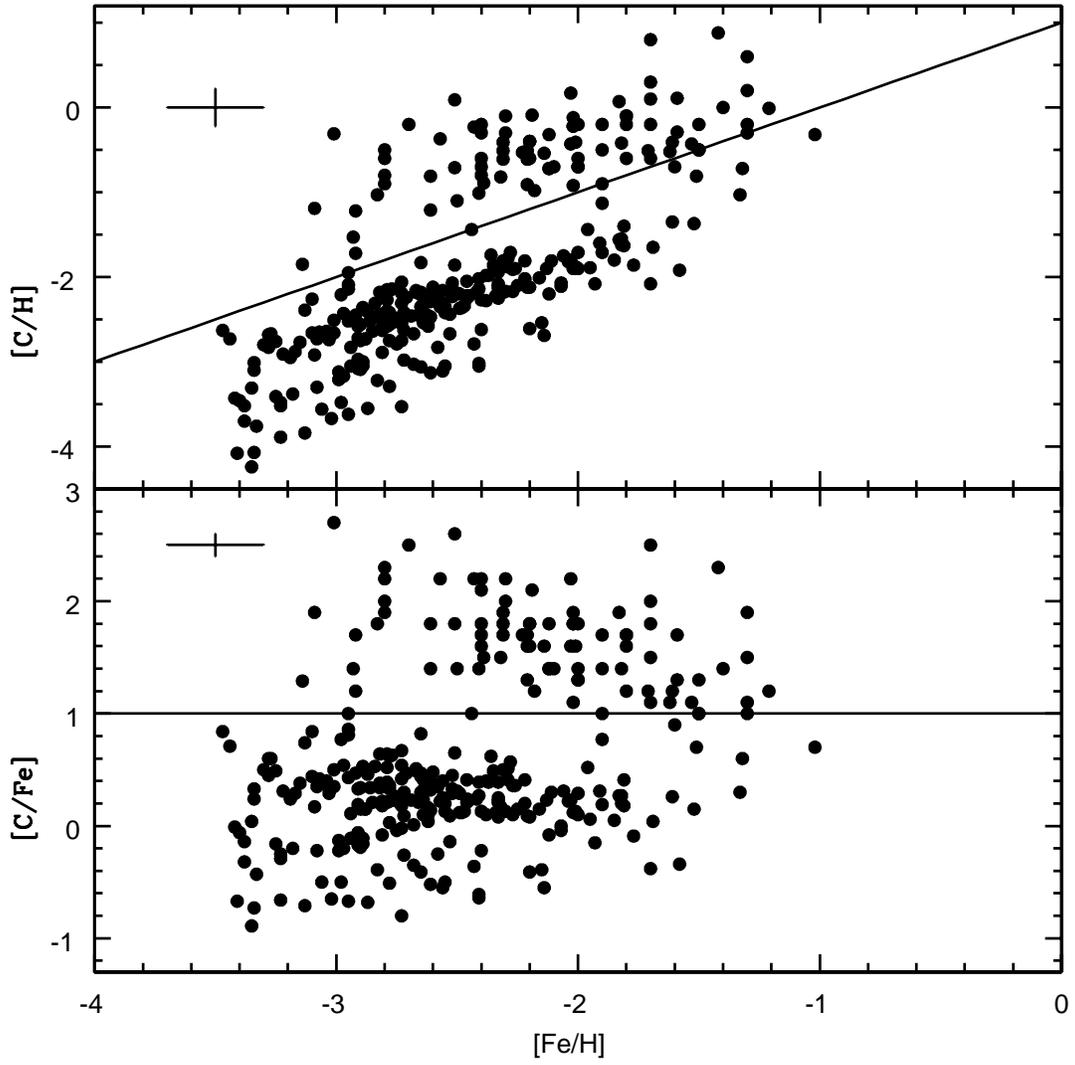}
\caption{[C/H] (top panel) and [C/Fe] (lower panel) vs. [Fe/H] for the total HERES sample.
The lines indicate our original cutoff for considering a star to be a CEMP star
([C/Fe] $\geq +1.0$). Typical error bars on the derived quantities are indicated.
\label{f_c_fe_he}}
\end{figure}

Including the HERES stars from Paper II and the additional stars analyzed
herein, we obtain a total of 270 HERES stars with [Fe/H] $\leq
-$2.0\footnote{Fe, C, and N abundances for the HERES C-rich sample are available
on request from the first author.}. Fifty-eight of these are CEMP stars, using
the definition [C/Fe] $\geq $+1.0. The observed fraction of CEMP stars among VMP
stars is thus 21$\pm$2\%. Figure
\ref{f_c_fe_he} shows the measured [C/H] and [C/Fe] as a function of [Fe/H] for
the full HERES sample.  Notice that quite a number of stars
lie only a small distance below the adopted cutoff. Inspection of 
the top panel of this Figure indicates that the upper limit to [C/H] appears
roughly constant at low [Fe/H], and approximately equal to the solar value.
As seen in the bottom panel of this Figure, it is also clear that 
the [C/Fe] ratio is rising with declining metallicity, as was previously noted 
for smaller samples of stars (e.g., Rossi et al. 2005). 
 
\begin{figure}
\plotone{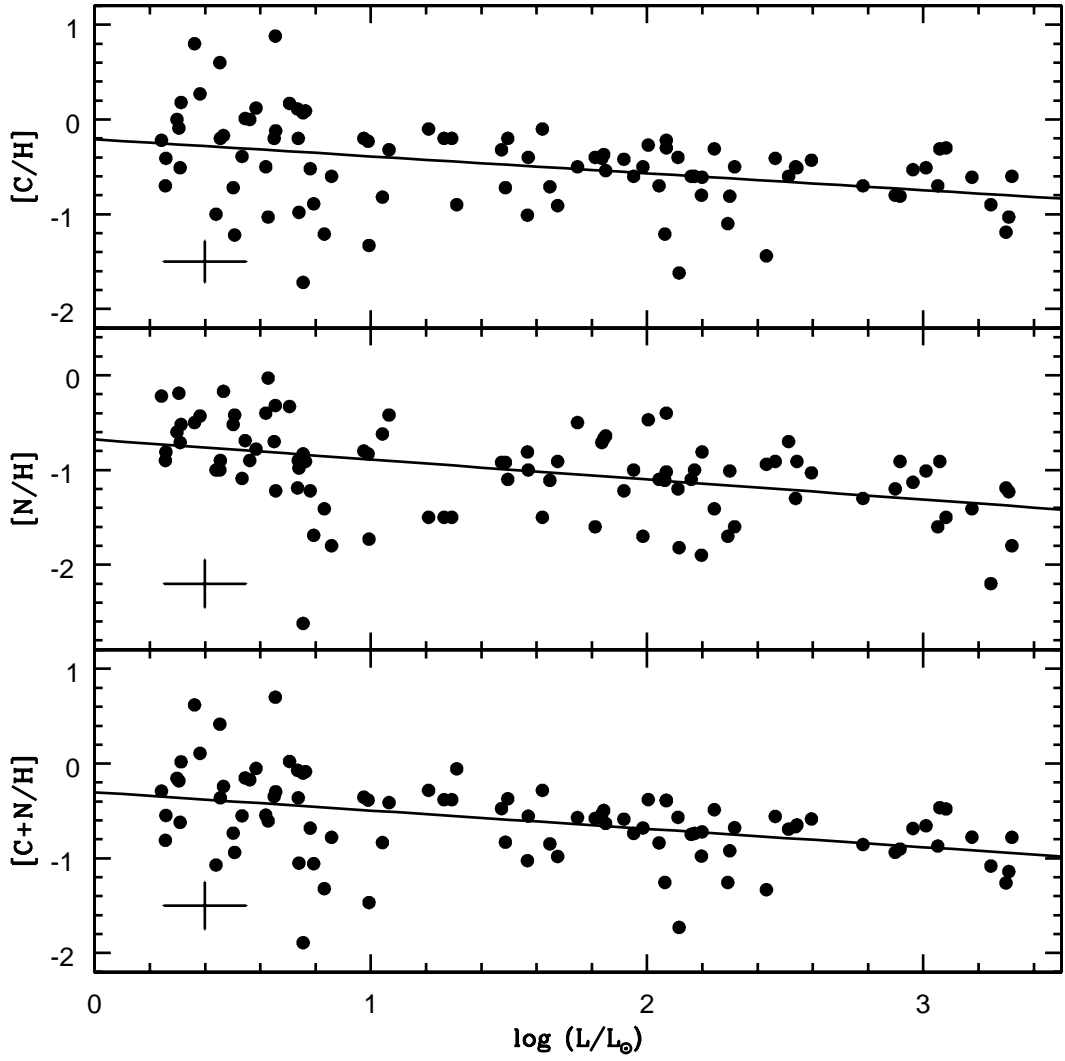}
\caption{[C/H] abundance (top panel), [N/H] abundance (middle panel),
 and [(C+N)/H] abundance (bottom panel), as a function of luminosity for the
HERES stars analyzed in the present paper.  The solid lines, with slopes $-0.18
\pm 0.05$ ([C/H]), $-0.21 \pm 0.05$ ([N/H]), and $-0.19 \pm 0.04$ ([(C+N)/H]), are linear 
regression fits to the data. Typical error bars on the derived quantities are indicated.
\label{f_ch_cnh_ev}}
\end{figure}

The HERES sample covers a wide range of luminosities, extending from below the
main-sequence turnoff up the red giant branch (RGB).  It is expected that the depth
of the stellar convective envelopes of these stars will vary considerably
with evolutionary state, with the cool giants having the possibility
of mixing a much larger amount of material than warmer dwarf and subgiant stars.
Figure \ref{f_ch_cnh_ev} shows the derived C, N, and C+N abundances as a
function of estimated luminosity. Inspection of this Figure reveals
that the points are not distributed uniformly; there are clear decreasing trends
of [C/H], [N/H], and [(C+N)/H] with luminosity.

As a star evolves off the main sequence, its convective envelope penetrates
deeper and deeper, reaching into regions where there was previous H burning, and
thus undergoes the so-called first dredge-up (at $\log $L/L$_\odot \sim$0.8). Later on, when
a star evolves off the RGB bump, a further mixing episode can take place (usually
referred to as ``extra mixing''). Both processes act in the direction of
decreasing C in favor of N.
 
If the stars under discussion had initially uniform chemical compositions (that
is, the composition of their outer envelopes reflected solely
that of the gas from which they were
born), their evolution would indeed affect the observed C and N abundances, but
in opposite directions. While the C abundance would drop, this would be
accompanied by an increase in N, due to CN processing in the interior of the
star; the [(C+N)/H] ratio should remain constant\footnote{Generally speaking,
the sum of C, N, and O is expected to remain constant in CNO processing. However
in the mass ranges under discussion ($\sim$ 1\,M$_\odot$) the envelope does not
reach deep enough to dredge up ON-processed material, and thus only C and N are
affected.}. This is indeed observed among metal-poor, C-normal field stars.
\citet{Spite06}, on the basis of a sample of 35 metal-poor stars, find that 
the abundances of N and C increase and decrease, respectively, in evolved stars 
with respect to unevolved ones. On the other hand, their sum remains constant 
\citep[see Fig 12 in][]{Spite06}.

However, if the C and N overabundances of the observed stars are limited to
their envelopes (as would arise from accretion of C,N-enriched material from the
ISM or from a companion), an additional effect should be taken into account. In
contrast to the previous case, when the convective envelope penetrates deeper,
it mixes in a considerable amount of originally unpolluted, pristine material.
Hence, the amount of C and N (as well as, when applicable, other elements for
which unusually high abundances are often found in CEMP stars, e.g., the
s-process elements) is diluted by mixing with a much larger H content, thus
decreasing {\it both} [C/H] and [N/H], as well as their sum. Although the above
mentioned mixing processes (first dredge-up and extra mixing) definitely occur
also in this case, their effect is expected to be much smaller than that due to
``evolutionary dilution,'' thus they would only contribute to the observed
scatter.

Although the observed correlations between [C/H], [N/H], and [(C+N)/H] and
luminosity shown in Figure~\ref{f_ch_cnh_ev} exhibit considerable scatter (the
correlation coefficients with luminosity being 0.46, 0.41, and 0.41 for [C/H],
[N/H], and [(C+N)/H], respectively), their presence support the role of
evolutionary dilution as the primary reason of the decrease in the observed C
(and N) abundances, and argues strongly in favor of an {\it external origin} for the
abundance patterns observed in the bulk of CEMP stars. 

\begin{figure}
\plotone{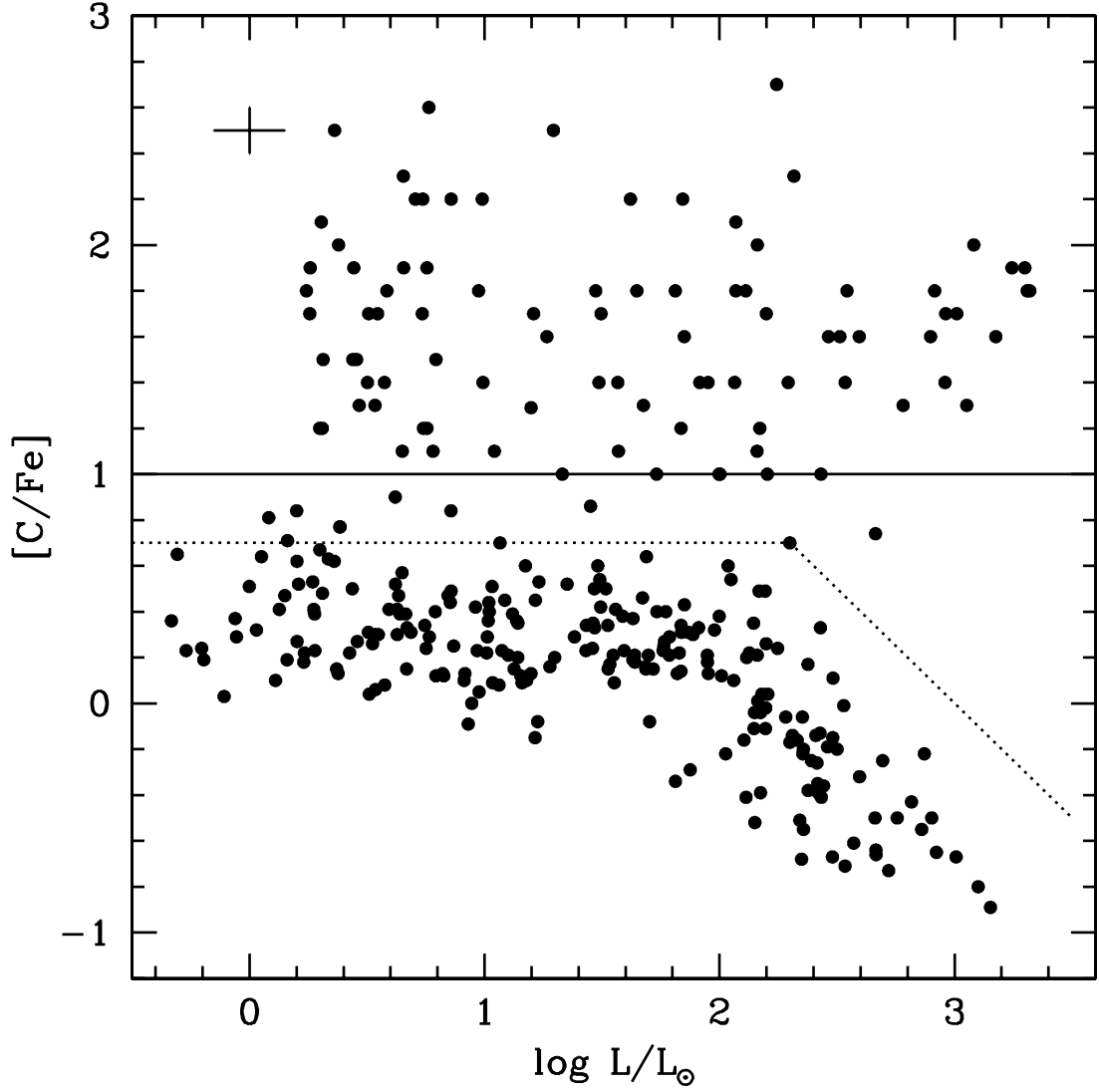}
\caption{[C/Fe] as a function of luminosity for the full HERES sample.
The solid line represents the cutoff for CEMP stars adopted in the 
present paper; the dotted line is that proposed by Aoki et al. (2006).
The dotted line has been calculated assuming a mass of 0.8\,M$_\odot$ for
all stars in the sample. A typical error bar on the derived quantities is indicated.
\label{f_c_feh_ev}}
\end{figure}

Due to the effects of evolutionary dilution, a number of stars that originally
had [C/Fe] $\geq +$1.0 at their main-sequence stage will end up with lower
observed [C/Fe] value in their more advanced evolutionary stages, and may not qualify as
CEMP stars according to our definition. Indeed, a CEMP fraction established on
the basis of a sample that includes cool giants is, strictly speaking, a lower
limit. One way of resolving this problem would be to adopt a
luminosity-dependent C-enhancement threshold. 
Aoki et al. (2006) have proposed
the following: a star is considered a CEMP star if it has [C/Fe] $\geq +$0.7 
and luminosity less than $\log(L/L_\odot)$ = 2.3.  For higher luminosity stars
the cutoff for classification as a CEMP star is [C/Fe]$\geq+3-\log(L/L_\odot)$.
Figure \ref{f_c_feh_ev} shows [C/Fe] as a function of luminosity for the
complete HERES sample. Also shown are two criteria for CEMP stars, [C/Fe] $\geq
+$1.0 for all evolutionary phases, and the Aoki et al. luminosity-dependent
criterion. If we adopt the Aoki et al. definition, a greater number of CEMP
stars are included, and the resulting fraction of CEMP stars in the HERES sample
rises to 24$\pm$3\%. However, most of the change in the fraction arises from 
the lowering of the cutoff in low-luminosity stars, rather than its 
luminosity dependence for evolved stars. The Aoki et al. criterion does not adequately 
account for possible alterations in the surface carbon abundances 
at luminosities below $\log(L/L_\odot)$ = 2.3, where, starting at luminosities 
as low as $\log(L/L_\odot)$ = 0.8, the first dredge-up occurs  and the 
convective envelope mass starts increasing.
Therefore, such criterion, even though definitely a step in the right direction, 
still does not fully take into account the evolutionary dilution effect described above.

Alternatively, it would be more meaningful to establish the frequency of CEMP stars based
exclusively on unevolved stars, in order to avoid the evolutionary
effects entirely. If we adopt $\log $(L/L$_\odot$)=0.8 as a working lower limit
for unevolved stars, the percentage of CEMP stars (using [C/Fe] = +1.0 as the
cutoff, independent of luminosity) among the VMP stars is 21$\pm$5\%. There are
only 68 HERES stars with [Fe/H]$ \leq -$2.0 and $\log$(L/L$_\odot)
\leq$0.8, therefore the inferred CEMP fraction is less certain.

\section{Conclusions}

From an analysis of the largest sample of VMP stars with available
high-resolution abundance determinations, the HERES sample, we find that the
fraction of CEMP stars with [Fe/H] $\leq -$2.0 is 21$\pm$2\%, confirming (some) previous
literature claims for a high frequency of CEMP stars at low metallicity.
Because of the evolutionary dilution effect discussed in this paper, this
fraction is very likely a lower limit. 
%Adoption of the luminosity-dependent
%criterion for the definition of CEMP stars proposed by Aoki et al. (2006) yields
%a value of 23$\pm$2\%. 

It should be kept in mind that the wide variety of elemental abundance
signatures observed in CEMP stars that have been analyzed at high spectral
resolution to date 
suggests that a number of different astrophysical processes are likely to be
involved in their production. Ideally, one would like to specify the frequency
of CEMP stars, {\it as a function of metallicity}, based exclusively on stars
with similar nucleosynthesis histories, rather than state only an aggregate
result below a fixed [Fe/H]. Knowledge of the dependency of CEMP frequency with
metallicity can shed light on the nucleosynthetic processes involved in the
production of C and N in the early Galaxy, as well as help to understand the
astrophysical formation sites and evolutionary histories of the CEMP stars.
%While insufficient data presently exist to populate the different [Fe/H] regimes
%(and CEMP categories), efforts are underway to procure this information.
%Marsteller et al. (in preparation) will report [Fe/H] and [C/Fe] for a large
%sample of unevolved CEMP stars obtained as part of SDSS-I (York et al. 2000). The Sloan
%Extension for Galactic Understanding and Exploration (SEGUE) survey is in the
%process of identifying many thousands of additional CEMP stars over a variety of
%evolutionary stages, including numerous unevolved stars. High-resolution
%spectroscopy of these stars will be obtained in due course, and is sure to
%provide interesting new constraints on the origin of CEMP stars.
\acknowledgements
S. L. acknowledges JINA for travel support and INAF COFIN 
for partial funding.
T.C.B., B.M., T.S., and Y.L. acknowledge partial funding for this work from grant
AST 04-06784, and PHY 02-16783: Physics Frontiers Center/Joint
Institute for Nuclear Astrophysics (JINA), both awarded by the U.S. National Science 
Foundation.  N.C. acknowledges financial support by Deutsche Forschungsgemeinschaft
through grants Ch~214/3 and Re~353/44. N.C.is a research fellow of the Royal Swedish 
Academy of Sciences supported by a grant from the Knut and Alice 
Wallenberg Foundation.  P.B. acknowledges support the Swedish 
Research Council.  S.R. acknowledges partial financial support from
the Brazilian institutions FAPESP, CNPq and Capes.

\clearpage

\clearpage

\clearpage

\end{document}